\documentclass[conference]{IEEEtran}
\IEEEoverridecommandlockouts

\usepackage[american]{babel}
\usepackage[utf8]{inputenc} 
\usepackage[backref=page]{hyperref}
\usepackage[scaled=0.92]{helvet}
\usepackage[nolist]{acronym}
\usepackage{amsfonts} 	% extra math fonts
\usepackage{amsmath} 	% mathematical formulas

\usepackage{amssymb} 	% Symbole für Zahlenmengen usw.
\usepackage{graphicx}
\usepackage{xcolor}
\usepackage{bookmark}
\usepackage{subcaption}
\usepackage{xspace}
\usepackage{tikz}
\usepackage{array}
\usepackage{multirow, makecell}
\usepackage{tabulary}
\usepackage{listings}
\usepackage{soul}
\usepackage{enumitem}

\acrodef{CG}{call graph}
\newcommand{\spa}{\textsc{SPA}\xspace}

\newcommand{\RQ}[1]{\textbf{RQ#1}\xspace} % research questions

\newcommand{\roundbox}[1]{
	\smallskip
	\noindent
	\begin{tikzpicture}
		\node[draw=black, rectangle, rounded corners](box){
			\begin{minipage}{\columnwidth}
				#1
			\end{minipage}
		};
	\end{tikzpicture}
}

\def\BibTeX{{\rm B\kern-.05em{\sc i\kern-.025em b}\kern-.08em
    T\kern-.1667em\lower.7ex\hbox{E}\kern-.125emX}}

\begin{document}

\title{How far are German companies in improving security through static program analysis tools?}

\author{\IEEEauthorblockN{Goran Piskachev, Stefan Dziwok, Thorsten Koch, Sven Merschjohan}
\IEEEauthorblockA{Fraunhofer IEM\\
\{name.surname\}@iem.fraunhofer.de}
\and
\IEEEauthorblockN{Eric Bodden}
\IEEEauthorblockA{Paderborn University 
	\& Fraunhofer IEM\\
eric.bodden@upb.de}
}

\maketitle

\begin{abstract}
As security becomes more relevant for many companies, the popularity of static program analysis (\spa) tools is increasing.  
In this paper, we target the use of \spa tools among companies in Germany with a focus on security. We give insights on the current issues and the developers' willingness to configure the tools to overcome these issues. 
Compared to previous studies, our study considers the companies' culture and processes for using \spa tools. 
We conducted an online survey with 256 responses and semi-structured interviews with 17 product owners and executives from multiple companies. 
Our results show a diversity in the usage of tools. 
Only half of our survey participants use \spa tools. 
The free tools tend to be more popular among software developers. 
In most companies, software developers are encouraged to use free tools, whereas commercial tools can be requested. However, the product owners and executives in our interviews reported that their developers do not request new tools. 
We also find out that automatic security checks with tools are rarely performed on each release.
\end{abstract}


\section{Introduction}
\label{sec:introduction}

In an increasingly digitalized world, the security of assets processed by software is highly relevant. One way to ensure better security is through static program analysis (\spa) tools.  The number of \spa tools with security focus is rising (e.g., CodeSonar~\cite{codesonar},  LGTM~\cite{lgtm}, Checkmarx~\cite{checkmarx}, Infer~\cite{infer}). From pattern-based approaches such as FindBugs~\cite{findbugs} to more complex data-flow analyses such as Checkmarx~\cite{checkmarx}, these tools have versatile applications with different stakeholders (e.g., software developers, testers, security experts). 

Due to their increasing popularity, usability researchers have studied \spa tools. Christakis et al.~\cite{christakis16} identified problems that software developers have while using \spa tools at \textit{Microsoft}. Lewis et al.~\cite{lewis13} performed user studies and interviews at \textit{Google} to investigate whether tools' findings influence the behavior of software developers. Nguyen Quang Do et al.~\cite{tse20nguyen} used a survey to identify the needs and motivators that software developers at \textit{Software AG} have for \spa tools. 

These studies mainly focus on software developers as primary users. Moreover, they are based on a single company case. A possible threat is that the company's culture and processes influence the software developers' perception. Hence, it is essential to address this when similar studies are conducted. 

This paper presents an empirical study on \spa tools performed in 2019 among German-speaking software development teams in multiple companies of different sizes. The results give insights into how software development teams use and configure \spa tools. The objective of the study is three-fold: (1) understanding the current usage of \spa tools in the industry, (2) identifying how software developers use \spa tools and configure them, and (3) understanding the companies' culture and processes for using \spa tools.   

To increase the credibility and validity of our research results, we apply a triangulation method consisting of an online survey and semi-structured interviews. 
Through the online survey, we target a more significant number of participants, in particular 256. In contrast, we target a more focused group of 17 experts from different companies through the interviews, including product owners and executives. The study reached a broad range of companies. However, 59 responses (23\%) from the survey are from a single large company, which remains anonymous. In the rest of the paper, we use the anonymous alias \textit{ABC}. The results show the same trend from the single company vs. multiple companies with a few minor exceptions we identified. 

In general, the use of \spa tools among the participants in our survey is 51.8\%, of which the free tools are the most popular among developers. Aligning with several existing studies, our study confirms some of the results of other studies, however also identifies new ones. E.g., the high number of false warnings remain a problem. In contrast to previous studies, the participants in our study think that the existing \spa tools are fast enough. 

Most participants are willing to configure the rules of the tools, even though only a few have experience with that. The product owners and executives are willing to provide enough tools to enable secure software development, but there is a low demand from their software developers. Finally, most participants think that the existing processes for \spa tools are not clearly defined. Although aware of the importance for automatic security checks before each release, these are performed only by third of the teams of our participants. 

The main contributions of this paper are as follows:

\begin{itemize}
	\item An online survey design targeting members of software development teams
	\item Semi-structured interviews with 17 product owners and executives from six companies
	\item Recommendations for the next generation tools based on results from this empirical study
	\item The artifact from our study is publicly available as an open science framework project\footnote{https://osf.io/k37c9/} including all relevant documents used to perform the survey and the interviews, the raw and processed data, and the anonymized protocols from the interviews.
\end{itemize}

\section{Related Work}
\label{sec:related_work}

This section lists relevant studies that we grouped into usability-related (\ref{sec:cat1}) and security-related (\ref{sec:cat2}) studies. 

\subsection{Usability of \spa tools}\label{sec:cat1}

Christakis et al.~\cite{christakis16} performed a study at \textit{Microsoft} consisting of a survey and interviews with developers, and analyzed data from high-critical bug reports handled by on-call engineers. As a result, they identified usability issues and missing functionalities that program analysis tools should address, such as aiming for a lower false positive rates, suppressing warnings, better fit into the workflow, etc. Nguyen Quang Do et al.~\cite{tse20nguyen} performed a study with the German company \textit{Software AG} by conducting a survey with developers and collecting usage data from the \spa tool \textit{Checkmarx}~\cite{checkmarx}. They identified users' needs when using \textit{Checkmarx}, such as better explainability of the warnings, an easier configuration of the rules, integrated recommendation system, and collaboration options. Sadowski et al.~\cite{sadowski15} performed another similar study at \textit{Google} by evaluating the usage of \textit{Tricoder}, a  platform for integrating different analysis tools to scan the codebase at \textit{Google}. The goal of the study was requirements elicitation for the platform. Luo et al.~\cite{magpiecloud} performed a user study to evaluate the cloud-based integration of the \textit{CodeGuru Reviewer} tool at \textit{Amazon}. Compared to our study, all studies targeted a single company, in particular large corporations, and a single tool to reason about usability issues and potential new features of the tools. 

Vassallo et al.~\cite{vassallo18} gathered participants from multiple companies for a survey and interviews to study the context in which the \spa tools are used. Similarly, Thomas et al.~\cite{thomas16} performed a user study and interviews with participants from multiple companies to study the context of the \spa tools within the IDE, in particular, the interactive features that the IDE provides. The usage of \spa tools within the CI pipelines was studied by Zampetti et al.~\cite{zampetti17} by mining open-source projects. Compared to our study, none of these studies consider the companies' culture and processes for using \spa tools.

\subsection{Security-related studies}\label{sec:cat2}

Smith et al.~\cite{Smith2019} used the tool \textit{FindBugs} in a user study and asked the participants to explain their thoughts while using the tool to find and fix security vulnerabilities. Using a card-sorting method with the collected data, they identified how users interact with \spa tools. Another study by Smith et al.~\cite{soups20} evaluated four \spa tools using a walkthrough method and interviews. This study evaluated the user interfaces of the selected \spa tools. As a result, they identified issues and potential improvements for the tools. Finally, Witschey et al.~\cite{Witschey2015} performed a survey and identified factors that impact the adoption of \spa tools. All studies focus on a given usage scenario. Moreover, none of them considered the impact of the company's culture and processes. 

Lastly, a recent study by Weir et al.~\cite{securitypassion} implemented intervention activities in eight development teams to increase the security awareness and security standards of their software products. These activities included trainings and workshops with the teams over four months. This study addressed a broad range of security activities such as threat modeling, penetration tests, security management, \spa tools, etc. The authors reported that these intervention activities, also for teams without security experts, are helpful. 
We collected similar data showing that our participants find activities, such as trainings and the use of \spa tools, highly relevant.

\section{Methodology}
\label{sec:study}

We present our triangulation methodology~\cite{empiricalmethods}, consisting of an online survey and semi-structured interviews. 
Both techniques are used to gather data from different roles. In the following, we state the research questions this paper addresses and the motivation behind them. Section~\ref{sec:survey} and Section~\ref{sec:interviews} discuss the method applied for the survey and the interviews, respectively. 

\subsection{Research questions}\label{sec:rqs}

As seen in the previous section, the use of \spa tools has been researched in multiple studies, targeting a selected company or specific context of use (e.g., use in IDE, use of a specific tool, etc.). Since the \spa research community has been growing, many new approaches have been published~\cite{popl19spds, stateofart1, stateofart2, stateofart3}. 
Additionally, new tools appeared, especially in the security domain, e.g., LGTM~\cite{lgtm}, SemGrep~\cite{semgrep}, Snyk~\cite{snyk}, CodeGuru~\cite{codeguru}, MarianaTrench~\cite{mt}. Therefore, studying the current practical use of these tools is still relevant. In particular, we investigate the use of \spa tools in one country, i.e., Germany. As security gains more attention among German companies~\cite{hiscox}, our study targets the companies' specific cultures and processes. Understanding the context in which \spa tools are currently used, will help tool vendors develop new features and better tools. 

This study answers the following research questions:

\begin{itemize}[leftmargin=1cm]
	\item[\RQ{1}] To what extent are \spa tools used in practice among software development teams in Germany?
	\item[\RQ{2}] How are \spa tools used in practice, and what are the problems faced by the users? 
	\item[\RQ{3}] What are typical culture and processes for using \spa tools and other security checks in German companies?
\end{itemize}

Since the study targeted software development teams in Germany, the survey and the interviews were conducted in German. The participation in the study was voluntary and without any personal compensation. 

\subsection{Survey}\label{sec:survey}

\subsubsection{Population}

To understand the usage of \spa tools, we invited participants from all roles involved in software development, including developers, architects, product owners, and executives. The study focuses on companies from Germany because our research institutions have contacts and regularly conduct projects with German companies. We used three ways to gather participants. First, we used our direct contacts. Second, we created posts on our institution's social media channels and website. Third, the survey was promoted by the media of one of the leading publishing houses in Germany,  \textit{Heise}~\cite{heise}, which publishes technology-focused magazines and organizes events with an audience mainly from German companies. Additionally, the survey was promoted among the networks \textit{Bitkom}\cite{bitkom}, \textit{it's OWL}~\cite{itsowl}, and \textit{innozent OWL}~\cite{innocent}. 
In total, we received responses from 350 participants. We excluded all incomplete responses, answered in an unrealistically short time, or not from Germany. After this filtering, we gathered 256 responses, of which 204 developers. If we consider that there are roughly 900.000 software developers in Germany\footnote{https://www.daxx.com/de/blog/entwicklungstrends/anzahl-an-softwareentwicklern-deutschland-weltweit-usa}, we get a margin of error of only 7\%\footnote{https://www.surveymonkey.com/mp/margin-of-error-calculator/}, which has a confidence level of 95\%, making our study representative of the target population.

Of all participants, 47\% are from large companies with more than 1000 employees, and the rest are from small- and medium-sized companies (Q24\footnote{throughout the paper we denote QX, to refer to the relevant question with number X from the online survey}). The size of most of the teams (Q29) is 6-15 persons (56\%) and about one third are small teams 1-5 persons (32\%). 61\% of the participants have long experience of more than ten years in software development (Q28). Of all participants, 80\% have \textit{software developer} role, followed by 12\% \textit{executive}, 10\% \textit{other}, 8\% \textit{product owner}, 6\% \textit{project manager}, 5\% \textit{data protection officer}, and 4\% \textit{security analyst}, where multiple answers were allowed (Q27). The participants come from 37 sectors, such as automotive, electrical, chemical, insurance, transportation, etc. (Q25).

\subsubsection{Data collection} We conducted the survey using the online tool \textit{Survey Monkey}~\cite{surveymonkey}. The tool allowed us to use different links for the six companies we invited and one link shared publicly for other companies. Among all links, we received a considerable number of responses from one large company: 59 responses from 256. Hence, in our results, we compare the trends among multiple companies of mixed size and a single large company. The survey was open for six weeks during summer 2019. On average, the participants needed 25 minutes to complete the questionnaire, measured based on the session duration per participant collected by Survey Monkey. 

\subsubsection{Design}

We conducted the survey as part of a large research project~\cite{appsecurenrw}, which goal is to understand the security-related activities in software development among German companies. This paper only presents the results on \spa tools. 

We followed the guidelines for opinion surveys by Kitchenham et al.~\cite{Kitchenham.2008}. Initially, we conducted a literature search to identify relevant related work (Section~\ref{sec:related_work}). None of the existing studies provided a survey instrument (i.e., a questionnaire) that can be reused. Hence, we created a new questionnaire for a cross-sectional survey. Five researchers were involved in creating and selecting the questions in a top-down process, starting from the research questions and breaking them into more concrete questions. The questionnaire was reviewed by three more researchers not part of the project. We performed two internal tests with students from our research group to verify the clarity of the questions and measure the time needed to complete the questionnaire. After that, we performed three external tests with professionals from the industry. 

The questionnaire with 42 questions is available in our artifact. The relevant questions for the paper are in the Appendix.  

\subsection{Interviews}\label{sec:interviews}

\subsubsection{Population} We performed 17 interviews with product owners and executives. Four of them were our previously known contacts. The rest were selected through convenience sampling~\cite{sampling}. We invited several randomly chosen companies from our region and used the \textit{first come, first served} principle to conduct the interviews with persons that volunteered to participate. Seven interviewees were product owners, six were executives, and the remaining four had both roles. All experts had professional experience in software development.  

\subsubsection{Data collection} Each interview was performed by two researchers. In total, six researchers took the role of interviewer. In each interview, one researcher was the moderator asking the questions and the other researcher wrote a protocol and, in rare cases, asked questions. Additionally, an audio recording of all sessions was made. After the interviews, the recordings were automatically transcribed and used to extend the protocols created during the interview. On average, each interview took 45 minutes. The interviews were conducted during the second half of 2019. For the evaluation of the interviews, we used the codebook method~\cite{codebook}, where three researchers manually annotated all transcripts.

\subsubsection{Design} We applied a similar process as the survey (Section~\ref{sec:survey}) to design the questionnaire for the interviews. We created two versions, one for each role, product owner and executives, which differ only in a few questions. The experts who had both roles were asked all questions.  

\section{Results}
\label{sec:results}

This section presents the study results and answers our research questions. With N, we denote the number of responses collected for each question. Note many participants answered different sections of the survey due to the optional questions. Hence, in the following, we report the percentage and the absolute numbers. To find correlations between two questions, we use the Cramer v value~\cite{cramer} (where values are between 0 and 1, with values over 0.25 having strong correlation) and for statictical significance the p value. 

\subsection{Use of \spa tools (\RQ{1})}
\label{sec:usage}

Among the survey participants, there is a heterogeneous use of IDEs and programming languages. The top three used IDEs (Q12) are IntelliJ IDEA (60\%), Eclipse (53\%), and Visual Studio Code (36\%). The top three used programming languages (Q13) are Java (76\%), JavaScript/TypeScript (45\%), and SQL (34\%). There is a correlation that those developers who use Java also use IntelliJ IDEA (Q13-Q12, Cramer V=0.537 p=0, N=123) and develop web applications (Q13-Q30, Cramer V=0.385 p=0.0001, N=114).

In total, 114 (51.8\%) out of 220 responses said they use \spa tools. When asked which tools they use (Q16), there were 57 uniquely named tools, of which only four tools were named in at least ten responses, i.e., SonarQube (50 responses), Findbugs (19), Checkstyle (13), and OWASP Dependency-Check (10). These tools are freely available or at least have a free version. They mostly perform simple pattern-based matching techniques to detect issues. SonarQube additionally has a taint analysis as more complex and sophisticated analysis. Other tools with more sophisticated dataflow analyses that the survey participants mentioned are: Checkmarx, Fortify, Klockwork, and Coverity. Only two participants noted that they use internally developed tools. 

Furthermore, the participants were asked to prioritize, where the warnings from the tools should be reported. From the possible options: (1) within the IDE, (2) on an internal website, and (3) in the ticket system, 160 out of 239 responses (66.9\%) selected the option (1) with the highest priority.

To compare the results within a single company with multiple differently-sized companies, we extracted the results from the company \textit{ABC}. 27 out of 59 responses (45.8\%) said that they use \spa tools which is lower than the 51.8\% among all companies including \textit{ABC} or 54\% excluding \textit{ABC}.  

\roundbox{There is a diversity in the use of IDEs and programming languages. Moreover, only about half of the teams use \spa tools, of which the most popular are SonarQube, Findbugs, Checkstyle, and OWASP Dependency-Check.}

\subsection{Problems and Tools Configuration (\RQ{2})}
\label{sec:configuration}

In the following, we discuss the problems identified by the participants when using the tools as well as their perspective on configuring the tools. 

Previous studies from several years ago ~\cite{christakis16} have reported that the existing tools are not fast enough to be used in development time. The participants in our study perceive the current situation differently. Ninety-three out of 114 collected responses (82\%) think that the tools they use are fast enough (Q18.1). However, the number of false warnings from the \spa tools remains high (Q18.2). Sixty-five out of 114 participants (57\%) have claimed this. In particular, those who answered that the number of false warnings is high also said that they program in the C language (Q18.2-Q13, Cramer V=0.305 p=0.0205, N=112)).

Based on their expertise and the warnings from the tools, 69\% of the participants can confirm which warnings are true positives (79 out of 114 responses) (Q18.3). The participants think that the tools find real issues regularly (80 out of 114 responses) (Q18.4). Finally, 81\% (92 out of 114) of the participants said that the messages reported from the warnings help them fix the issues found in the code (Q18.5). 

When it comes to the configuration of the tools, we asked the participants to what extent they are willing to change, add, or remove the rules used by the tools to find different security issues. Sixty-one percent (69 out of 114 responses) are willing to define their own custom rules to be used by the tool (Q18.6). Nevertheless, only 35\% (40 out of 114 responses) have experience defining custom rules for the tools. From them, many have answered that they have the role of \textit{Security Analyst} (Cramer V=0.377 p=0.0026, N=108). Those participants who are willing to define custom rules are in teams that have testing responsibilities (Q18.8-Q2, Cramer V=0.313 p=0.0163, N=114). Eighty-three percent (95 out of 114 responses) of the participants are willing to provide feedback to the tools in terms of marking false positives to get better results in future runs of the tools (Q18.7).  

When observing the results from \textit{ABC}, there is a slight higher willingness with 63\% (17 out of 27 responses) to configure or with 89\% (24 out of 27 responses) to provide feedback to the tools. For the questions on the quality of the results, there are only minor differences (under 3 \%) except for one: 78\% (21 out of 27 responses) of the participants from \textit{ABC} think that the tools find real issues regularly, compared to 68\% (59 out of 87 responses) to the rest. 

Finally, we asked the participants in the interviews if they allow their software developers to invest time in providing feedback to the \spa tools. Fourteen participants answered, of which only one did not agree. The reasoning behind not allowing this is that there is a risk of providing feedback, due to lack of expertise, may reduce the true warnings. 

\roundbox{Our results show that users consider static code analysis tools fast enough but still have issues with the high number of false warnings. The messages of the warnings are helpful for most of the users to fix the issues. Most users are willing to invest time in adapting the rules of the tools or provide feedback for improved future results.}

\subsection{Culture and processes for using \spa tools (\RQ{3})}
\label{sec:regulations}

The most popular tools listed by the participants from Section~\ref{sec:usage} are free tools. The most popular commercial tools listed are CheckMarx and Fortify, listed only 5 and 4 times, correspondently (Q17). We asked the product owners and executives about their opinion on open and free tools. Fourteen have answered this question, of which all allow their software developers to use free and open source \spa tools. Even most of them encourage the teams to use open and free \spa tools. When asked whether there is a budget for commercial \spa tools, only one participant said that there is no budget. Ten participants said there is only a budget when there are requests, whereas six said there is a dedicated budget for this purpose. One participant commented \textit{"These tools are a good investment"}. Most interviewees said that developers rarely request commercial tools. 

\roundbox{Even though there is a budget for \spa tools in most German companies that we interviewed, the free \spa tools are still more popular among the developers than the commercial tools.}

The participants from \textit{ABC} have different options than the rest of the participants regarding the availability of tools for secure software development. Fifty-nine percent (16 out of 27 responses) from \textit{ABC} think that they have the right amount of tools, whereas only 38\% (33 out of 87 responses) have this opinion in the other companies. 

In particular, our study targeted the security aspect of the processes. We asked for each phase (requirements, design, and implementation/testing) and whether the security is considered during the activities. Security is considered differently in each phase: 57\% in requirements (Q4.1), 76\% in design (Q6.1), and 75\% in implementation/testing (Q10.1). When we asked all participants if they use tools to check the security properties of their code (Q10.3), 52\% answered positive (59 out of 114 responses) and 43\% negative (49 out of 114 responses). 
Moreover, among the participants that use \spa tools, only 17\% (11 out of 63 responses) answered that they perform an official security review before each release (Q22.1), while 80\% do not perform and 3\% do not know. Similarly, only 36\% (38 out of 134 responses) said that they perform automatic security checks after the release is built (Q22.2), while 62\% disagree and 2\% do not know. In addition, only 22\% said that they perform automatic security checks on the software while operating (Q22.3), while 68\% disagree and 10\% do not know. 

There is a correlation showing that the teams, in which the developer checks the security requirements, are in need of better tools to accomplish their tasks (Q11-Q15.2, Cramer V=0.364 p=0.0444, N=67). There is also a correlation showing that when the security requirements are checked, these checks are done during development time (Q14-Q10.3, Cramer V=0.376 p=0.00007, N=123). The participants who answered that they perform security checks before the release also have a dedicated security team (Q22.2-Q11, Cramer V=0.455 p=0.000001, N=123) and/or hire external companies for security (Q22.2-Q11, Cramer V=0.358 p=0.0001, N=123).

Finally, only 29\% of the participants (28 out of 98 responses) have a clear process how to verify the implementation concerning security (Q10.4), whereas 64\% do not (63 out of 98 responses). Nine percent of the participants (21 out of 221 responses) said that nobody is responsible for checking the the software's specified security requirements (Q11). For 34\% (75 out of 221 responses), the programmer does this. Only for 16\% (36 out of 221 responses), this is performed by a security team, or 9\% (20 out of 221 responses) from an external company.

\roundbox{Concerning the security processes, many companies do not have clear regulations. The executives see value in having tools and are willing to allocate enough budget.}

\subsection{Discussion}
\label{sec:discussion}

In the following, we discuss the findings from our study, including few anecdotes. 

\paragraph{Open-source tools vs. commercial tools} In \RQ{1}, we found out that software developers more use the free \spa tools. Additionally, via \RQ{3}, we found that executives are willing to provide a (more) budget for commercial \spa tools. These findings are contrast to some degree, leaving space for discussion and speculation. One possible explanation might be that the executives have higher expectations of the tools and their developers when a costly tool is requested. Hence, developers are more comfortable using free and open-source \spa tools. One of the executives in our interviews said: \textit{"They are using more and more tools because they do not cost anything. They use beautiful graphics, which in many places leads me to the fact that the relevant thing is that these numbers are there, but interpretations does not take place."}. The same executive has further elaborated the expectations about commercial tools: \textit{"So, if you now put your feet on the table for 6,000 euros, then at some point someone will stand in front of the door... You wanted the 6,000 euros, now show me what you're doing with it! Most of the time, no one asks that with free tools. They then say (for the free tools) 'Oh how nice is that, yes it's orange. At least it's not red."}.

Moreover, some executives also have high expectations from the commercial tools when it comes to the stability of their processes. One of the product owners in our interviews said: \textit{"A paid tool naturally has the charm that you can hang the responsibility around the neck of tool manufacturers."}. Later, referring to the tool manufacturers, he said: \textit{"We'll pay them. If anything goes wrong, it's their fault... It's a relatively convenient measure for the managers"}. As seen in this statement, the executive who invests in commercial \spa tools is shifting the responsibility to the tool manufacturers. 

\paragraph{Configuring the tools} Concerning \RQ{2}, we asked the participants about the effort they are willing to put into improving the tools. This includes activities such as labeling the findings or configuring the rules used by the tool i.e., writing new rules or changing the existing rules. In the interviews, one executive said that the interpretation of the results is much more important for him than improving the tools. 
While some executives are not convinced that putting effort into improving the tools is a good investment of time, few others stated the opposite. Most participants in the study think that providing feedback to the tool is a good idea.  
Studies have shown that the configuration of the tools requires codebase-specific information~\cite{issta19swan}.  

\paragraph{Fast \spa tools} Via \RQ{2}, we found out that the participants consider the \spa tools fast enough. In contrast, previous studies~\cite{christakis16, bradley16} reported the opposite. A possible explanation for this is that the most popular tools named among our participants are free \spa tools that perform more lightweight analyses (pattern-based) compared to most commercial tools with deep data-flow analyses.  

\paragraph{\spa tools integrations} Developers use the \spa tools in different workflows, e.g., as part of the integrated development environment (IDE), integrated pipeline builds, or in external apps such as dashboards. Most executives reported that they favor having the tools as part of the pipeline, with other tests, as a continuous integration step, as one said, \textit{"because this is done continuously, permanently anyway"}. However, the developers want fast feedback as they need to work on issues and improve the code. Most participants in the study, most of whom are developers, said that the results should be reported in the IDE. They want the issues to be reported as early as possible and not wait for a possible slow build pipeline. Our recommendation for tool manufacturers is to develop integrations for both views, as many tools on the market already do.

\paragraph{False warnings} The issue of the high number of false warnings from the \spa tools have been reported in multiple previous studies. Many of the participants in our interviews have confirmed this issue as well. Since there is a high interest from the survey participants in providing feedback to the tools and reconfiguring the rules of the tools, the issue of "false positives" can be avoided when users are given more time and appropriate training on how to use the tools more effectively. Recent studies from the static analysis community have shown the importance of the configuration of the tools concerning the results~\cite{analyzers}. However, as earlier stated, some executives have slight scepticism. This requires further research. One possible factor is the required security expertise of the developers who are using the tools. This can be confirmed by the fact that only 69\% of the participants reported that they can confirm which warning is a true positive (Question 17.3). The rest of the participant need appropriate security training. 

\section{Threats to Validity}
\label{sec:threats}

We conducted a survey study based on the guidelines by Kitchenham et al.~\cite{KPP95} and Runeson et al.~\cite{RHRR12}. Next, we discuss construct validity, external validity, and reliability. 
\paragraph{Construct Validity}

The set of questions used in the survey and the interviews is the outcome of several workshops with software security researchers. A possible threat is the level of expertise that these persons have which influences the quality of the questions. Moreover, we avoided the possibility that the interviewers ask wrong questions as we - the researchers that conducted this study - held the interviews ourselves. 
Moreover, we knew only four of the 17 interviewed persons upfront - thus, they were not influenced by us. We pre-tested our survey and interviews with people from our target group to identify whether our questions were understandable and interpreted as intended.

\paragraph{External Validity}

As our study was conducted only with companies from Germany, it might be the case that our results do not apply to companies outside of Germany as they have other standards and laws. % that they have to consider. 
Though several international standards like ISO IEC 27001 exist. 

Even though we made  pre-tests, it might be the case that the survey participants misinterpreted some questions. In our semi-structured interviews, this possibility is even less as our interviewees always had the chance to ask whether they understood our questions correctly. Our survey is representative for our intended target group. However, the number of interviews might not be sufficient for a representative result.

\paragraph{Reliability}

The questions of our study are based on our experience. Three of the authors are in the late stage of their Ph.D. with +6 years of academic experience, and the other two authors are on a senoir level with a proven academic record. To minimize human errors, we used as many automated tools when possible, e.g., we used Survey Monkey to collect the survey data, and we recorded all interviews and used a reliable voice-to-text software for the transcription. With a self-created script, we processed all raw data wherever possible.

\section{Ethics}
\label{sec:ethics}

The participation in the study was voluntary. The participants in the interviews signed a consent form. For most questions, we provided an option for participants not wanting to give any details (i.e., “I don’t know”). Furthermore, we aligned our study to the data protection laws in Germany and the EU. The questions were reviewed at our institution by multiple researchers, including one expert on professional trainings and surveys, the head of the department, the data protection officer, and one of the directors. 

\section{Conclusion}
\label{sec:conclusion}

We presented our study among software development teams in differently-sized companies in Germany. We conducted an online survey reaching a large number of software developers and other roles involved in software development and 17 semi-structured interviews with product owners and executives. 

Our study shows that \spa tools are used in different contexts and only by half of our participants. It confirms that there are still issues such as the high number of false warnings, but there are also improvements, such as the analysis runtime. The participants are willing to configure the tools and provide feedback to better results. In the future similar studies should be conducted in other countries to compare the situation in different regions and cultures. In Germany, further studies should be performed to understand why commercial tools are not requested by the development teams even though executives are willing to allocate a budget. 
Future research should help us understand whether the existing trainings on security and \spa tools are sufficient, or new ones are needed. 

\section*{Acknowledgement} We acknowledge the funding by the project "AppSecure.nrw - Security-by-Design of Java-based Applications" of the European Regional Development Fund (ERDF-0801379). We thank Katharina Altemeier for her contribution to the questionnaires and to Sebastian Leuer and Boris Budweg for their contribution in processing the data. 

\bibliographystyle{ieeetr}
\bibliography{bibliography}

\newpage
\appendix

\section{Survey Questions}\label{appendix:survey}
The survey consists of 7 parts. We list all questions from the survey that are relevant for this paper. 

\textbf{Part 1}: Questions for all roles
\begin{itemize}
	\item[1] To what extent do you agree with the following statements? [likert scale: "strongly agree", "agree", "disagree", "strongly disagree", "do not know"]
	\begin{itemize}
		\item[1.1] For the topic \textit{Secure Software Engineering} (SSE) our team invests the right amount of time.
		\item[1.2] In our team, we have members who are responsible for \textit{Security}. 
		\item[1.3] We have clearly defined regulations and policies how to develop secure software. 
		\item[1.4] We have the right amount of tools to support us in developing secure software. 
		\item[1.5] Our development process and the existing tools are enough for our needs. 
		\item[1.6] Security requirements (e.g. secure processing of confidential data) are clearly defined in our team.  
		%\item[1.7] Each member in our team should have high competences in the topic SSE.
	\end{itemize}
	\item[2] What is your team responsible for regarding your software product? [multiple choice: "Deployment/Server configuration", "Programming", "Testing", "Build processes", "Software product operation", "Requirements", "Architecture and design"]
\end{itemize}

\textbf{Part 2}: Questions related requirements
\begin{itemize}
	\item[3] Is your role involved in requirements elicitation? [single choice: "Yes", "No" (if no, then part 2 is skipped)] 
	\item[4] Please answer the following questions [single choice: "Yes", "No", "I don't know"]
	\begin{itemize}
		\item[4.1] Is security considered during activities related to requirements management?
		\item[4.2] Do you have security experts that check the security requirements?
	\end{itemize}
	\item[5] To what extent do you agree with the following statement related to security requirements? [likert scale: "strongly agree", "agree", "disagree", "strongly disagree", "do not know"]
	\begin{itemize}
		\item[5.1] Our current processes should be more precise and clear.
		\item[5.2] More or better tools would help us to perform our tasks with higher quality. 
	\end{itemize}
\end{itemize}

\textbf{Part 3}: Questions related design and architecture
\begin{itemize}
	\item[6] Is your role involved in design and architecture? [single choice: "Yes", "No" (if no, then part 3 is skipped)] 
	\item[7] Please answer the following questions [single choice: "Yes", "No", "I don't know"]
	\begin{itemize}
		\item[7.1] Is security considered during activities related to design and architecture?
		\item[7.2] Do you have a process to check the security properties of the design with respect to the implemented program?
		\item[7.3] Do you have security experts that check the architecture from security perspective?
	\end{itemize}
	\item[8] To what extent do you agree with the following statement related to secure design and architecture? [likert scale: "strongly agree", "agree", "disagree", "strongly disagree", "do not know"]
	\begin{itemize}
		\item[8.1] Our current processes should be more precise and clear.
		\item[8.2] More or better tools would help us to perform our tasks with higher quality. 
	\end{itemize}
\end{itemize}

\textbf{Part 4}: Questions related to implementation and testing
\begin{itemize}
	\item[9] Is your role involved in implementation and testing? [single choice: "Yes", "No" (if no, then part 4 is skipped)]
	\item[10] Please answer the following questions [single choice: "Yes", "No", "I don't know"]
	\begin{itemize}
		\item[10.1] Is security considered during implementation?
		\item[10.2] Are there templates, in particular standards for implementing secure software?
		\item[10.3] Do you use tools to automatically check security checks of the implemented code?
		\item[10.4] Do you have a process to check the security properties of the code?
	\end{itemize}
	\item[11] Who checks the code agains security vulnerabilities? [multiple choice with option for free text: "Same person who wrote the code", "Another person from the team who did not write the code", "Internal security team", "External security team", "Nobody", "Other"]
	\item[12] Which IDEs are used within your team? [multiple choice with option for free text: "Eclipse", "IntelliJ Idea", "NetBeans", "Visual Studio Code", "Visual Studio", "vi / vim", "Notepad++ (or similar editor)", "Apple Xcode", "Other"]
	\item[13] Which programming languages are used primarily within your team? [multiple choice with option for free text: "Java", "JavaScript/TypeScript", "C\#", "C", "C++", "Kotlin", "Objective-C", "Python", "Ruby", "PHP", "Go", "SQL", "Swift", "Rust", "R", "Other"]
	\item[14] When is the program checked against security vulnerabilities? [multiple choice with option for free text: "During implementation in the IDE", "Before each commit in the repository", "After each commit from the server pipeline", "Before official release", "During one sprint", "Never", "Other"]
	\item[15] To what extent do you agree with the following statement related to secure software implementation and testing? [likert scale: "strongly agree", "agree", "disagree", "strongly disagree", "do not know"]
	\begin{itemize}
		\item[15.1] Our current processes should be more precise and clear.
		\item[15.2] More or better tools would help us to perform our tasks with higher quality. 
	\end{itemize}
\end{itemize}

\textbf{Part 5}: Questions related to \spa tools
\begin{itemize}
	\item[16] Does your team use static analysis tools, such as \spa tools? [single choice: "Yes", "No" (if no, then part 5 is skipped)]
	\item[17] Which \spa tools are used within your team? [open question]
	\item[18] To what extent do you agree with the following statements? [likert scale: "strongly agree", "agree", "disagree", "strongly disagree", "do not know"]
	\begin{itemize}
		\item[18.1] The tools we use return the results fast enough for our needs.
		\item[18.2] The number of reported warnings that are false (false positives) is too high. 
		\item[18.3] I can confirm the true warnings (true positives) easily. 
		\item[18.4] The tools we use often report true warnings. 
		\item[18.5] The messages of the warnings help me to fix the issues in the code. 
		\item[18.6] I am willing to write our project-specific custom rules for the \spa tools. 
		\item[18.7] I am willing to label the false warnings to give feedback to the tools so that the tools can improve in the future. 
		\item[18.8] I have experience in writing custom rules for some of the tools. 
	\end{itemize}
	\item[19] Please sort the following statements based on importance, where 1 has the highest importance. [sorting of statements from 1 to 4]
	\begin{itemize}
		\item The analysis should finish within few seconds.
		\item The messages of the reported findings should be understandable and provide hints how to fix the issues. 
		\item It should be easy for me to understand and adapt the rules of the tools according to my needs.
		\item The tool should return only very few false warnings. 
	\end{itemize}
	\item[20] Where should the warnings from the \spa tools be reported? [sorting of statements from 1 to 3]
	\begin{itemize}
		\item In my IDE (e.g. Eclipse)
		\item On an internal website (e.g. Jenkins) 
		\item In our ticket system (e.g. Jira)
	\end{itemize}
\end{itemize}

\textbf{Part 6}: Questions related to software operation and maintenance
\begin{itemize}
	\item[21] Is your role involved in software operation and maintenance? [single choice: "Yes", "No" (if no, then part 6 is skipped)]
	\item[22] Please answer the following questions [single choice: "Yes", "No", "I don't know"]
	\begin{itemize}
		\item[22.1] Does your team performs a final security review before each release?
		\item[22.2] Are there automatic security checks for each release?
		\item[22.3] Are there automatic security checks during operation?
	\end{itemize}
	\item[23] To what extent do you agree with the following statements related to secure software operation and maintenance? [likert scale: "strongly agree", "agree", "disagree", "strongly disagree", "do not know"]
	\begin{itemize}
		\item[23.1] Our current processes should be more precise and clear.
		\item[23.2] More or better tools would help us to perform our tasks with higher quality. 
	\end{itemize}
\end{itemize}

\textbf{Part 7}: Meta-data questions 
\begin{itemize}
	\item[24] How many employees has your company? [single choice: "1-3", "4-10", "11-50", "51-250", "251-1000", "$>$ 1000"] 
	\item[25] In which domain operate your company? [open question]
	\item[26] How many employees work in software development? [single choice: "1-50", "51-250", "$>$ 250"]
	\item[27] What is your position? [multiple choice with option for free text: "Management", "Project lead", "Product owner", "Software development (requirements, implementation, testing)", "Security analyst", "Information security officer", "Other"] 
	\item[28] How many years of experience in software development do you have? [single choice: "$<$ 2 years", "2-5 years", "6-10 years", "$>$ 10 years"] 
	\item[29] How many members has your team? [single choice: "1-5", "6-15", "16-30" "$>$ 30"] % YES !
	\item[30] What type of applications do you develop? [multiple choice with option for free text: "Mobile", "Desktop", "Web", "Embedded", "Server", "Other"]
\end{itemize}

\end{document}